\RequirePackage{silence}
\documentclass[fleqn,usenatbib,useAMS]{mnras} 
\usepackage{graphicx}
\usepackage{amsmath}

\usepackage{amsfonts}
\usepackage{float}
\usepackage{bm}
\usepackage[perpage,symbol*]{footmisc}
\usepackage{ae,aecompl}
\usepackage{array}
\usepackage{soul}
\usepackage{mathtools}
\usepackage{multirow}

\newcommand\azero{a_{_0}}
\newcommand\gzero{g_{_0}}

\newcommand\gNzero{g_{_{N0}}}
\newcommand\vg{\bm{g}}
\newcommand\vgzero{\vg_{_0}}
\newcommand\vgN{\bm{g}_{_N}}
\newcommand\vgNzero{\vg_{_{N0}}}
\newcommand\dvg{\bar{\bm{g}}}
\newcommand\dvgN{\dvg_{_N}}
\newcommand\vk{\bm{k}}
\newcommand\vr{\bm{r}}
\newcommand\PhiN{\Phi_{_N}}
\newcommand\gNr{g_{_{Nr}}}
\newcommand\gNz{g_{_{Nz}}}
\newcommand\grA{g^{^A}_{_r}}
\newcommand\hw{\widehat{\bm{w}}}

\title[Toomre stability of disk galaxies in QUMOND]{Toomre stability of disk galaxies in quasi-linear MOND} 

\author[Indranil Banik, Mordehai Milgrom and Hongsheng Zhao]{Indranil Banik$^{1, 2}$\thanks{Email:
\href{mailto:ibanik@astro.uni-bonn.de}{ibanik@astro.uni-bonn.de} (Indranil Banik)
\newline $~~~~~~~~~~~~~~~$
\href{mailto:moti.milgrom@weizmann.ac.il}{moti.milgrom@weizmann.ac.il} (Mordehai Milgrom)
\newline $~~~~~~~~~~~~~~~$
\href{mailto:hz4@st-andrews.ac.uk}{hz4@st-andrews.ac.uk} (Hongsheng Zhao)}, Mordehai Milgrom$^{3}$ and Hongsheng Zhao$^{1}$\\
$^{1}$Scottish Universities Physics Alliance, University of Saint Andrews, North Haugh, Saint Andrews, Fife, KY16 9SS, UK\\
$^{2}$Helmholtz-Institut f\"ur Strahlen und Kernphysik (HISKP), University of Bonn, Nussallee 14$-$16, D-53115 Bonn, Germany\\
$^{3}$Department of Particle Physics and Astrophysics, Weizmann Institute, Rehovot 7610001, Israel}

\pubyear{2018}
\begin{document}
\label{firstpage}
\pagerange{\pageref{firstpage}--\pageref{lastpage}}

\maketitle

\begin{abstract}

We consider disk stability in the quasi-linear formulation of MOND (QUMOND), the basis for some $N$-body integrators. We generalize the Toomre criterion for the stability of disks to tightly wound, axisymmetric perturbations. We apply this to a family of thin exponential disks with different central surface densities. By numerically calculating their QUMOND rotation curves, we obtain the minimum radial velocity dispersion required for stability against local self-gravitating collapse.

MOND correctly predicts much higher rotation speeds in low surface brightness galaxies (LSBs) than does Newtonian dynamics without dark matter. Newtonian models thus require putative very massive halos, whose inert nature implies they would strongly stabilize the disk.

MOND also increases the stability of galactic disks, but in contradistinction to Newtonian gravity, this extra stability is limited to a factor of 2. MOND is thus rather more conducive to the formation of bars and spiral arms. Therefore, observation of such features in LSBs could be problematic for Newtonian galaxy models. This could constitute a crucial discriminating test. We quantitatively account for these facts in QUMOND.

We also compare numerical QUMOND rotation curves of thin exponential disks to those predicted by two algebraic expressions commonly used to calculate MOND rotation curves. For the choice that best approximates QUMOND, we find the circular velocities agree to within 1.5\% beyond $\approx 0.5$ disk scale lengths, regardless of the central surface density. The other expression can underestimate the rotational speed by up to 12.5\% at one scale length, though rather less so at larger radii.


\end{abstract}


\section{Introduction}
\label{Introduction}

MOND \citep{Milgrom_1983} is an alternative to the dark matter (DM) hypothesis in accounting for the observed dynamical discrepancies in galactic systems, especially between their measured rotation curves (RCs) and those predicted by Newtonian gravity \citep[e.g.][]{Rubin_Ford_1970, Rogstad_1972}. A discrepancy is also apparent in the `timing argument': in Newtonian gravity, the visible masses of the Galaxy and M31 are insufficient to turn their initial post-Big Bang recession around to the observed extent \citep{Kahn_Woltjer_1959}. MOND posits that these acceleration discrepancies\footnote{sometimes called mass discrepancies, though the reason for the discrepancy may not be missing mass} are not due to the presence of DM but arise from a breakdown of Newtonian dynamics that is reckoned without in analyzing the dynamics of these systems. MOND is extensively reviewed in \citet{Famaey_McGaugh_2012} and \citet{Milgrom_2014}. The M31 timing argument in MOND was discussed by \citet{Zhao_2013} and detailed calculations were presented in section 2 of \citet{Banik_Ryan_2018}.

MOND introduces $\azero$ as a fundamental acceleration scale of nature. When the gravitational field strength $g \gg \azero$, standard dynamics is restored. However, when $g \ll \azero$, the dynamical equations become scale-invariant \citep{Milgrom_2009_DML}. In this deep-MOND regime, an inevitable consequence of scale invariance is that, asymptotically far outside a distribution of total mass $M$, the rotational speed of a test particle becomes independent of its distance $R$ from the mass. This occurs when $R \gg r_{_M}\equiv \sqrt{GM/\azero}$, where $r_{_M}$ is the MOND radius of the mass $M$. Therefore, in a modified gravity formulation of MOND, $g\propto R^{-1}$ beyond the MOND radius. Applying dimensional arguments to the fact that $\azero$ is the only additional constant in MOND, we must have that
\begin{eqnarray}
	g ~ \propto ~ \frac{\sqrt{GM\azero}}{R} ~~~\text{for } ~ R ~\gg ~ r_{_M}\equiv \sqrt{\frac{GM}{\azero}} \, .
	\label{Deep_MOND_limit}
\end{eqnarray}
The normalization of $\azero$ is taken so that the proportionality here becomes an equality. Empirically, $\azero \approx 1.2 \times {10}^{-10}$ m/s$^2$ to match galaxy RCs \citep[e.g.][]{McGaugh_2011, Li_2018}. Note that MOND gravity must be non-linear in the matter distribution because $g \propto \sqrt{M}$.

It was noticed early on \citep[e.g.][]{Milgrom_1983, Milgrom_1999} that, remarkably, this value is similar to accelerations of cosmological significance. For example,
\begin{eqnarray}
	2 \mathrm{\pi} \azero ~\approx ~ a_{_H} \left( 0 \right) ~\equiv ~ c H_0 ~\approx ~ a_{_\Lambda} ~\equiv ~ \frac{c^2}{\ell_\Lambda} \, ,
\end{eqnarray}
where $H_0$ is the present-day value of the Hubble constant and $\ell_\Lambda = \left( \Lambda/3 \right)^{-1/2}$ is the de Sitter radius corresponding to the observed value of the cosmological constant $\Lambda$ \citep[e.g.][]{DES_results_2018}.

Stated another way, $\azero$ is similar to the acceleration at which the classical energy density in a gravitational field \citep[][equation 9]{Peters_1981} becomes comparable to the dark energy density $u_{_\Lambda} \equiv \rho_{_\Lambda} c^2$ implied by $\Lambda$.
\begin{eqnarray}
	\frac{g^2}{8\mathrm{\pi}G} ~<~ u_{_\Lambda} ~~\Leftrightarrow~~ g ~\la~ 2\mathrm{\pi}a_{_0} \, .
	\label{MOND_quantum_link}
\end{eqnarray}
This association of local MOND with cosmology suggests that MOND may arise from quantum gravity effects \citep[e.g.][]{Milgrom_1999, Pazy_2013, Verlinde_2016, Smolin_2017}.

Regardless of its underlying microphysical explanation, MOND correctly predicted the RCs of a wide variety of both spiral and elliptical galaxies across a vast range in mass, surface brightness and gas fraction \citep[e.g.][]{Lelli_2017, Li_2018}. Although internal accelerations are harder to measure within ellipticals, this can sometimes be done accurately when they have a surrounding $X$-ray emitting gas envelope in hydrostatic equilibrium \citep{Milgrom_2012} or a thin rotation-supported gas disk \citep[e.g.][]{Heijer_2015, Serra_2016}. The success of MOND extends down to pressure-supported galaxies as faint as the satellites of M31 \citep{McGaugh_2013} and the Milky Way, though in the latter case one must be careful to exclude galaxies where MOND predicts significant tidal distortion \citep{McGaugh_2010}. For a recent overview of how well MOND works in several different types of galaxy across the Hubble sequence, we refer the reader to \citet{Lelli_2017}.


It is worth emphasizing that MOND does all this based solely on the distribution of luminous matter. It is clear that these achievements are successful a priori predictions because most of these RCs $-$ and sometimes even the baryon distributions $-$ were measured in the decades after the first MOND field equation was put forth \citep{Bekenstein_Milgrom_1984} and its new fundamental constant $\azero$ was determined \citep{Begeman_1991}. These predictions work due to underlying regularities in galaxy RCs that are difficult to reconcile with the collisionless DM halos of the $\Lambda$CDM paradigm \citep[e.g.][]{Salucci_2017, Desmond_2016, Desmond_2017}.

These halos were originally introduced to boost the RCs of disk galaxies. If real, they would also endow their embedded disks with added stability because the halo contribution to the gravitational field responds very little to density perturbations in the disk, making the disk more like a set of test particles. This fact forms the basis of another argument originally adduced for the presence of DM halos around disk galaxies \citep{Ostriker_Peebles_1973} $-$ without such halos, observed galactic disks would be deleteriously unstable \citep{Hohl_1971}.

A crucial prediction of MOND was that low surface brightness galaxies (LSBs) would show large acceleration discrepancies at all radii because $g \ll \azero$ everywhere in the galaxy \citep{Milgrom_1983}. This prediction was thoroughly vindicated by later observations \citep[e.g.][]{Blok_1997, McGaugh_1998}.

In $\Lambda$CDM, such LSBs must be assigned a halo much more massive than the disk. Such a halo would cause the disk to become very stable, stymieing the formation of bars and spiral arms which are believed to result from disk instabilities \citep{Lin_1964}.

Since MOND posits that dark halos are absent, the question arises regarding the degree of stability of disks, especially LSB disks in which MOND is predicted to have significant effects. This issue was discussed in some detail by \citet{Milgrom_1989}, who showed that MOND generally does add to the stability of low-acceleration disks with a given mass distribution and velocity dispersion. The reason is that instead of the Newtonian relation $g_d\propto \Sigma$ between surface mass density $\Sigma$ and the acceleration $g_d$ it produces in the disk, the deep-MOND relation is $g_d \propto \sqrt{\Sigma}$. This means that in Newtonian disks without a DM halo, density perturbations $\delta \Sigma$ produce acceleration perturbations
\begin{eqnarray}
	\frac{\delta g_d}{g_d} ~\sim ~ \frac{\delta\Sigma}{\Sigma} \, .
	\label{Delta_Ln_g}
\end{eqnarray}
Adding a non-responsive DM halo with contribution $g_h$, we can write $g_d = g_b+g_h$, where $g_b$ is the Newtonian contribution of the baryonic disk that satisfies Equation \ref{Delta_Ln_g}. Because density perturbations in the disk do not affect $g_h$, we have that
\begin{eqnarray}
	\frac{\delta g_d}{g_d} ~\sim ~ \frac{\delta \Sigma}{\Sigma} \left( 1 + \frac{g_h}{g_b} \right)^{-1}.
\end{eqnarray}
The reduced response of $g_d$ implies increased disk stability.

The analogous result in the deep-MOND regime is
\begin{eqnarray}
	\frac{\delta g_d}{g_d} ~\sim ~ \frac{1}{2} \frac{\delta \Sigma}{\Sigma}
\end{eqnarray}
because $g_d \propto \sqrt{\Sigma}$ and $g_h = 0$. The added degree of stability in deep MOND is thus similar to that endowed by a halo with $g_h \sim g_b$. As MOND effects are strongest in this regime, it is clear that this is the limit to how much MOND enhances the stability of disk galaxies, even those with very low surface densities. However, the massive halos required by $\Lambda$CDM would increase the stability indefinitely as the surface density is reduced. This makes deep-MOND LSBs develop spiral arms and fast-rotating bars more readily than according to $\Lambda$CDM \citep{Tiret_2007, Tiret_2008}.

Beyond such general semi-qualitative arguments, it is important to study disk stability more quantitatively in specific, action-based MOND theories. This not only gives a more accurate picture, it is also necessary for understanding and avoiding the development of instabilities in numerical codes based on these theories.

\citet{Milgrom_1989} studied disk stability analytically in the context of the aquadratic Lagrangian formulation of MOND \citep[AQUAL,][]{Bekenstein_Milgrom_1984}. This was followed by the numerical studies of \citet{Brada_1999}, which demonstrated disk stability in MOND by solving the AQUAL field equation using an expensive non-linear grid relaxation stage. This is unavoidable in the context of AQUAL and remains an aspect of more recent codes that solve it \citep{Londrillo_2006, Candlish_2015}.
	
Since that time, another non-relativistic, action-based MOND theory has been put forth. This quasi-linear formulation of MOND \citep[QUMOND,][]{QUMOND} is much more amenable to numerical simulations because the grid relaxation stage is linear, like in Newtonian gravity. Despite using different field equations to implement MOND consistently, QUMOND and AQUAL give rather similar results, as demonstrated both numerically \citep{Candlish_2016} and analytically \citep{Banik_2015}.

QUMOND can be solved numerically using the publicly available $N$-body and hydrodynamics code Phantom of RAMSES \citep{PoR}, an adaptation of the grid-based RAMSES algorithm widely used by astronomers due to its adaptive mesh refinement feature \citep{Teyssier_2002}.\footnote{The similar code RAyMOND can also handle AQUAL \citep{Candlish_2015}. It will eventually be made public.} As a result, QUMOND has become the main workhorse for simulations of galaxy evolution and interactions \citep{Thies_2016, Renaud_2016, Thomas_2017, Thomas_2018, Bilek_2018}.

The question of disk stability in QUMOND remains to be addressed analytically despite its importance for understanding the results of such simulations and for establishing their initial conditions. Here, we study some aspects of disk stability in QUMOND. We focus on deriving an analytic expression for the QUMOND generalization of the Toomre criterion \citep{Toomre_1964}. This gives an estimate of whether part of a disk is vulnerable to self-gravitating collapse. In particular, the Toomre criterion gives a lower limit to the radial velocity dispersion in a thin disk for it to remain stable against short-wavelength axisymmetric perturbations.

After outlining the broader context in Section \ref{Introduction}, we explain the steps in our analytic derivation (Section \ref{Section_QUMOND}). Our main result is a generalization of the Toomre condition to QUMOND (Equation \ref{Min_Q_star_QUMOND} for stellar disks and Equation \ref{Min_Q_gas_QUMOND} for gas disks). In Section \ref{Numerical_results}, we apply the former to numerically determined RCs of a family of thin exponential disks with different central surface density. Observational constraints on disk stability are discussed in Section \ref{Observations}. We conclude in Section \ref{Conclusions}.

\section{Stability analysis in QUMOND}
\label{Section_QUMOND}

QUMOND \citep{QUMOND} is a modified gravity formulation of non-relativistic MOND that is derivable from an action principle. It is also the non-relativistic limit of a certain bimetric theory of relativistic MOND \citep{Milgrom_2009}. In QUMOND, the gravitational potential $\Phi$ is determined (given suitable boundary conditions) by
\begin{eqnarray}
	\nabla^2\Phi ~=~ \nabla \cdot \left[ \nu \left( \frac{ \left| \nabla \Phi_N \right|}{\azero} \right) \nabla \Phi_N \right],
	\label{QUMOND_basic}
\end{eqnarray}
where $\nu(y)$ is the so-called MOND interpolating function (in the $\nu$ version) between the Newtonian and modified regimes. Our results in this section do not depend on the specific choice of $\nu$ function, but one observationally motivated form is given in Equation \ref{Simple_nu}. To recover Equation \ref{Deep_MOND_limit}, $\nu$ must satisfy the limits $\nu \left( y \right) \to 1$ for $y \gg 1$ (Newtonian regime) and $\nu \left( y \right) \to \frac{1}{\sqrt{y}}$ for $y \ll 1$ (deep-MOND regime). $\Phi_N$ is an auxiliary potential that solves the unmodified Poisson equation with the mass density $\rho$ as its source. It is the Newtonian potential of the system, but this in itself plays only an indirect role in the dynamics.

In this section, we determine the stability condition of a thin, flat axisymmetric QUMOND disk to tightly wound Wentzel-Kramers-Brillouin (WKB) axisymmetric perturbations. For this purpose, we consider perturbations $\dvg$ to the acceleration field $-\nabla \Phi \equiv \vg \equiv \vgzero +\dvg$, where $\vgzero \equiv -\nabla \Phi_0$ is the unperturbed or background field ($_0$ subscripts indicate unperturbed quantities while ${\, \bar{} \,}$ superscripts denote perturbations). In these terms, Equation \ref{QUMOND_basic} reads
\begin{eqnarray}
	\nabla \cdot \left( \vgzero  + \dvg \right) ~=~ \nabla \cdot \left[ \nu \left( \frac{\left|  \vgNzero+\dvgN \right|}{\azero} \right) \left(\vgNzero+ \dvgN \right) \right].
	\label{QUMOND_governing_equation}
\end{eqnarray}

The $_N$ subscripts refer to Newtonian quantities, which must first be calculated in order to solve Equation \ref{QUMOND_governing_equation}. Since perturbations in the auxiliary intermediate Newtonian field drive those in the actual QUMOND field, we first discuss how $\vg_{_N} \equiv -\nabla \PhiN$ is affected by a density perturbation.

\subsection{Perturbation to the Newtonian potential}
\label{Delta_g_N}

The Toomre criterion (which our work generalizes) concerns a localized (wavelength ${\ll r}$) axisymmetric perturbation to the Newtonian potential $\PhiN$ \citep{Toomre_1964}. This perturbation satisfies the Laplace equation outside the disk because there is no matter there. In cylindrical polar co-ordinates $\left(r, z \right)$ chosen so $z = 0$ is the disk mid-plane, this becomes
\begin{eqnarray}
	\frac{1}{r}\frac{\partial}{\partial r}\left(r\frac{\partial \bar{\Phi}_{_N}}{\partial r}\right) \, + \, \frac{\partial^2 \bar{\Phi}_{_N}}{\partial z^2} ~=~ 0 \, .
	\label{Laplace_equation}
\end{eqnarray}

In the WKB approximation, $\bar{\Phi}_{_N}$ changes on a scale $\ll r$, reducing this to
\begin{eqnarray}
	\frac{\partial^2 \bar{\Phi}_{_N}}{\partial r^2} \, + \, \frac{\partial^2 \bar{\Phi}_{_N}}{\partial z^2} ~=~ 0 ~~\text{(WKB) }.
	\label{Laplace_equation_WKB}
\end{eqnarray}

The WKB assumption greatly simplifies our analysis, which would otherwise have to consider geometric corrections in addition to variation of the background surface density within a single perturbation wavelength. Nonetheless, use of the WKB approximation is an important shortcoming of our work, which we discuss further in Section \ref{Section_WKB}. Its main result is that our analysis breaks down sufficiently close to the centre of a galaxy (Figure \ref{L_crit}).

To analyze the stability of a disk to perturbations of different wavelengths, we assume the surface density $\Sigma$ is perturbed with real radial wave-vector $\alpha$ such that the density perturbation is the real part of
\begin{eqnarray}
	\bar{\Sigma} ~=~ \widetilde{\Sigma} \, \mathrm{e}^{\mathrm{i}\alpha r}.
	\label{Delta_Sigma}
\end{eqnarray}

We expect that the resulting perturbation to $\PhiN$ takes the form
\begin{eqnarray}
	\bar{\Phi}_{_N} ~=~ \widetilde{\Phi}_{_N} \,\mathrm{e}^{\mathrm{i}\vk \cdot \vr} \, ,
	\label{Phi_N_tilde}
\end{eqnarray}
where $\vk \equiv \left( k_r, k_z \right)$ is independent of the position $\vr$ relative to the disk centre.

Given the nature of the density perturbation $\bar{\Sigma}$ (Equation \ref{Delta_Sigma}), we assume $k_r = \alpha$. Outside the disk, Equation \ref{Laplace_equation_WKB} implies that ${k_r}^2 + {k_z}^2 = 0$, so $k_z$ must be imaginary. Choosing the sign so that $\bar{\Phi}_{_N}$ decays away from the disk on both sides and equating the discontinuity in $\frac{\partial \bar{\Phi}_{_N}}{\partial z}$ with $4 \pi G \, \bar{\Sigma}$, we get that
\begin{eqnarray}
	\bar{\Phi}_{_N} ~=~ -\frac{2\mathrm{\pi}G \widetilde{\Sigma}}{\left| \alpha \right|} \mathrm{e}^{\left(\mathrm{i} \alpha r - \alpha \left| z \right| \right)} \, .
	\label{Newtonian_solution}
\end{eqnarray}

In what follows, we consider the region $z > 0$. Due to reflection symmetry about the disk mid-plane, the results apply to both sides if we make the identification $z \to \left| z \right|$. We also assume that $\alpha > 0$ as local disturbances with wave-vectors $\pm \alpha$ should behave in the same way given the equality between the real parts of $\mathrm{e}^{\pm \mathrm{i}\alpha r}$.

\subsection{Linearized QUMOND}

For a WKB stability analysis, we consider small perturbations $\dvg$ whose wavelength is much shorter than the scale over which the background $\vgzero \left( \vr \right)$ varies (typically the maximum of $\left| \vr \right|$ and the disk scale length). We use the governing Equation \ref{QUMOND_governing_equation} on the side $z > 0$ in the region outside the disk. In this region, $\vgzero$ varies only a little over many perturbation wavelengths, and as usual in our Toomre-type analysis it can be assumed constant. As the background fields satisfy Equation \ref{QUMOND_basic}, we focus on the first order perturbative terms in Equation \ref{QUMOND_governing_equation}.
\begin{eqnarray}
	\label{QUMOND_g_ext_domination}
	\nabla \cdot \dvg ~&=&~ \nu_{_0} \left[ \nabla \cdot \dvgN + K_0 (\hw\cdot\nabla) {\bar g}_{_{Nw}}\right], ~\text{where}\\
	\nu_{_0} ~&\equiv&~ \nu \left( \frac{\gNzero}{\azero} \right) ~~\text{and} \\
	K_0 ~&\equiv&~ \left.\frac{d \ln[\nu(y)]}{d \ln(y)}\right|_{y=\gNzero/\azero}.
\end{eqnarray}

Here, $\gNzero \equiv |\vgNzero|$ and $\hw$ is a unit vector in the direction of ${-\vgzero}$. This lets us define ${\bar g}_{_{Nw}}\equiv \dvgN \cdot \hw$ i.e. ${\bar g}_{_{Nw}}$ is the component of $\dvgN$ in the $\hw$ direction. Equation \ref{QUMOND_g_ext_domination} can be obtained by considering the QUMOND dynamics of a system whose internal accelerations are much smaller than the system's acceleration in a constant `external' field, represented here by $\vgzero$ on one side of the disk. Such an external field-dominated situation was treated in equation 67 of \citet{QUMOND} and in equation 24 of \citet{Banik_2015}, with the latter work explaining the derivation. In this quasi-Newtonian limit, the $\nu$ function is linear, making MOND gravity linear in the matter distribution.

\subsection{Perturbation to the QUMOND potential}
\label{Section_QUMOND_potential}

Substituting the Newtonian potential perturbation from Equation \ref{Newtonian_solution} into Equation \ref{QUMOND_g_ext_domination} gives the equation governing the QUMOND potential perturbation.
\begin{eqnarray}
	\label{QUMOND_forcing}
	\nabla^2 \bar{\Phi} ~&=&~ Q \mathrm{e}^{\mathrm{i} \alpha r - \alpha z}, \\
	Q ~&\equiv&~ 2 \mathrm{\pi} G \nu_{_0}K_0 \widetilde{\Sigma} \alpha \left(\cos 2\theta + \mathrm{i} \sin 2\theta \right),
\end{eqnarray}
where $\theta$ is the angle $\hw$ makes with the outwards radial direction just above the disk plane. In a realistic astrophysical disk, $\vgNzero$ is partly directed towards lower $r$ such that $\theta \leq \frac{\pi}{2}$.

As is well known, there are two parts to the general solution of an inhomogeneous equation like Equation \ref{QUMOND_forcing}. The first is an arbitrary multiple of the complementary function $\Phi_c$. In this case, $\Phi_c$ corresponds to solving the homogeneous Laplace equation with $\mathrm{e}^{\mathrm{i} \alpha r}$ dependence at ${z = 0}$. We already showed in Section \ref {Delta_g_N} that
\begin{eqnarray}
	\Phi_c ~=~ A \mathrm{e}^{\mathrm{i} \alpha r - \alpha z},
\end{eqnarray}
for any constant $A$.

The second part of the solution is the particular integral $\Phi_p$ of the inhomogeneous Equation \ref{QUMOND_forcing}. The particular integral of Equation \ref{QUMOND_forcing} must also have the same $\mathrm{e}^{\mathrm{i} \alpha r}$ dependence at $z = 0$. To exploit our knowledge of $\Phi_c$, we try a solution of the form $\Phi_p = f(z)\mathrm{e}^{\mathrm{i} \alpha r - \alpha z}$. Substituting this into Equation \ref{QUMOND_forcing}, we obtain that $f'' - 2 \alpha f' = Q$, where $'$ indicates a radial derivative. Of the two resulting constants of integration, one is absorbed by adjustments to $A$. The other is fixed by requiring $\bar{\Phi}$ to decay away from the disk plane, implying that $\Phi_p$ must do so. Thus, we obtain that
\begin{eqnarray}
	\Phi_p ~=~ -\frac{Qz}{2\alpha} \mathrm{e}^{\left(\mathrm{i} \alpha r - \alpha z \right)} \, .
\end{eqnarray}

Noting that our derivation can be extended to both sides of the disk by replacing $z \to \left| z \right|$, we see that the general solution is
\begin{eqnarray}
	\bar{\Phi} = \mathrm{e}^{\left(\mathrm{i} \alpha r - \alpha \left| z \right| \right)} \left[A - \mathrm{\pi} \left| z \right| \nu_{_0} K_0 G \widetilde{\Sigma}\left(\cos 2\theta + \mathrm{i} \sin 2\theta \right) \right] \, .
	\label{Phi_A_QUMOND}
\end{eqnarray}

Similarly to the Newtonian case (Equation \ref{Newtonian_solution}), $\bar{\Phi}$ decays exponentially with $ \left| z \right|$ over a perturbation wavelength, though the decay is of the form $z \mathrm{e}^{-z}$ rather than $\mathrm{e}^{-z}$. A more striking difference is that the QUMOND $\bar{\Phi}$ and $\bar{\Sigma}$ are not in phase as $\sin 2\theta \neq 0$ in general. For a disk stability analysis, only $\bar{\Phi}$ at ${z = 0}$ is relevant and this does have the same phase as $\bar{\Sigma}$. The phase difference becomes apparent only when ${z \neq 0}$.

In Newtonian gravity, such a phase offset is impossible because the problem is symmetric with respect to ${r \to -r}$ at minima and maxima in $\bar{\Sigma}$. This is not so in QUMOND because it depends non-linearly on the total gravitational field, \emph{including its background value}. This generally has some radial component ($\cos \theta \neq 0$), thus picking out a preferred direction and breaking the reflection symmetry. This unavoidable effect in MOND causes it to violate the strong equivalence principle, a phenomenon known as the external field effect \citep{Milgrom_1986}.

\subsection{The boundary condition}
\label{Boundary_condition}

In this section, we fix the constant $A$ in Equation \ref{Phi_A_QUMOND} using the boundary condition on $\bar{\Phi}$ at the surface of the disk. For this purpose, we apply Gauss' theorem to the linearized QUMOND equation with constant $\vgzero$ (Equation \ref{QUMOND_g_ext_domination}). This tells us that, at $z=0^+$, we must have equality between the $z$ components of the vectors (indicated with $_z$ subscripts) under the divergence on both sides of this equation.
\begin{eqnarray}
	{\bar g}_{_z} ~=~ \nu_{_0} \left({\bar g}_{_{Nz}} + K_0 \, {\hw}_{_z}{\bar g}_{_{Nw}}\right) \, .
\end{eqnarray}

Upon differentiating the Newtonian potential from Equation \ref{Newtonian_solution}, we get that
\begin{eqnarray}
	{\bar g}_{_{Nw}} ~=~ 2\mathrm{\pi} G \widetilde{\Sigma} \left( -\mathrm{i} \sin \theta + \cos \theta \right)\mathrm{e}^{\mathrm{i} \alpha r} \, .
\end{eqnarray}

As a result, the boundary condition becomes
\begin{eqnarray}
	&&-A\alpha - \nu_{_0}\mathrm{\pi}GK_0\widetilde{\Sigma} \left( \cos^2 \theta - \sin^2 \theta + 2\mathrm{i} \sin \theta \cos \theta \right) \nonumber \\
	&&= ~ 2\mathrm{\pi}G\nu_{_0}\widetilde{\Sigma}\left[1 + K_0 \sin \theta \left(-\mathrm{i} \cos \theta + \sin \theta \right) \right] \, .
	\label{Gaussian_integral_condition_QUMOND}
\end{eqnarray}

This can be simplified to give
\begin{eqnarray}
	A ~=~ -\frac{\mathrm{\pi} G\nu_{_0}\widetilde{\Sigma}\left( 2 + K_0 \right)}{\left| \alpha \right|} \, .
	\label{MOND_Phi_amplitude}
\end{eqnarray}

In the Newtonian limit where $K_0 = 0$ and $\nu_{_0} = 1$, this reduces to $A = -\frac{2\mathrm{\pi}G\widetilde{\Sigma}}{\left| \alpha \right|}$, thereby reproducing Equation \ref{Newtonian_solution} and equation 5.161 of \citet{Galactic_Dynamics}.

\subsection{The final result}

The response of the disk to a density perturbation is governed by the resulting potential perturbation within the disk plane. Restricting our more general result for $\bar{\Phi}$ (Equation \ref{Phi_A_QUMOND}) to ${z = 0}$ shows that disk stability is governed by the parameter $A$. Comparing its value in Equation \ref{MOND_Phi_amplitude} with what it would be in Newtonian gravity, we get that the effective $G$ entering a local stability analysis is boosted by the factor
\begin{eqnarray}
	\gamma ~=~ \nu_{_0}\left( 1 + \frac{K_0}{2} \right).
	\label{gamma}
\end{eqnarray}

Thus, the disk stability results in \citet{Galactic_Dynamics} can be generalized to QUMOND if we multiply its equation 5.161 by $\gamma$ and follow the change through. This is because $G$ does not affect the restoring force arising from velocity dispersion (analogous to pressure in a gas). Once the background $\vgzero $ is fixed, so is the RC. In this case, $G$ also has no effect on the restoring force from disk shear. Only the self-gravitating term \citep[e.g.][equation 6.66]{Galactic_Dynamics} is affected, so $G$ only enters the stability criterion inasmuch as it affects the relation between $\widetilde{\Sigma}$ and $A$. Therefore, the Toomre stability condition for Newtonian disks \citep{Toomre_1964} can be generalized to QUMOND by setting
\begin{eqnarray}
	G ~\to~ G \nu_{_0} \left( 1 + \frac{K_0}{2} \right).
	\label{Toomre_QUMOND}
\end{eqnarray}

\citet{Milgrom_1989} derived the corresponding result for AQUAL, whose governing field equation is
 $\nabla \cdot \left[ \mu(|\vg|/\azero) \vg \right] = \nabla \cdot \vgN$. In this theory, the analogue of Equation \ref{Toomre_QUMOND} is
\begin{eqnarray}
	\label{Toomre_AQUAL}
	G &\to& \frac{G}{\mu_{_0} \sqrt{1 + L_0}}, ~\text{ where} \\
	\mu_{_0} &\equiv& \mu \left( \frac{\gzero}{\azero} \right) ~\text{ and} \\
	L_0 &\equiv& \left. \frac{d \ln \left[ \mu \left( x \right) \right]}{d \ln \left( x \right)} \right|_{x= \gzero/\azero}~~.
\end{eqnarray}

The results in AQUAL and QUMOND are rather similar $-$ when we approximate each theory by its algebraic analogue (exactly valid in spherical symmetry), QUMOND reduces to $\vg = \nu\left( \frac{\left| \vgN \right|}{\azero} \right) \vgN$ and AQUAL to $\mu \left( \frac{\left| \vg \right |}{\azero} \right) \vg = \vgN$. In this one-dimensional situation, the theories are equivalent provided $\nu \left( \frac{\left| \vgN \right|}{\azero} \right) \mu \left( \frac{\left| \vg \right|}{\azero} \right) = 1$. Assuming this to be the case, \citet{Banik_2018_Centauri} showed in their section 7.2 that this pair of $\vg$ and $\vgN$ satisfy
\begin{eqnarray}
	\left( 1 + K_0 \right) \left( 1 + L_0 \right) ~=~ 1.
	\label{L0_K0_relation}
\end{eqnarray}

To first order in $K_0$ or $L_0$, this implies equality between the factors of $\nu_{_0} \left( 1 + \frac{K_0}{2} \right)$ in Equation \ref{Toomre_QUMOND} and $\frac{1}{\mu_{_0} \sqrt{1 + L_0}}$ in Equation \ref{Toomre_AQUAL}. The disk stability condition is therefore very similar in both formulations of MOND. This suggests that simulations of disk galaxies in the more computer-friendly QUMOND should yield rather similar results to AQUAL \citep{Candlish_2015}.

In general, Equation \ref{L0_K0_relation} is not exactly correct because the algebraic relations are only approximations. However, we show in Section \ref{Section_ALM} that the QUMOND version is very accurate beyond the central scale length of a thin exponential disk galaxy. Similar results were previously obtained for AQUAL \citep{Brada_1995}.

\subsection{Towards non-axisymmetric perturbations}
\label{Non_axisymmetric_perturbations}

Our analysis so far has focused exclusively on short-wavelength axisymmetric perturbations. Observed spiral galaxies often have spiral arms with a near-constant pitch angle \citep{Seigar_1998}. These logarithmic spirals were considered in detail by \citet{Kalnajs_1971}, who obtained several Newtonian analytic results regarding their stability.

Locally, a logarithmic spiral is similar to the plane waves assumed in our WKB analysis, but with one important difference $-$ the wave vector is not parallel to the outwards radial direction. Instead, it lies at some angle $\psi$ to this direction. In Newtonian gravity, this should not affect the resulting perturbation to $\PhiN$ because, in a local analysis, there is nothing special distinguishing the radial and azimuthal directions. Thus, Equation \ref{Laplace_equation_WKB} remains unaltered if $r$ is replaced with $\widetilde{r}$, a co-ordinate pointing along the wave-vector rather than the outwards radial direction. The co-ordinate system $\left( \widetilde{r}, z \right)$ is still orthogonal because the wave vector is still within the disk plane.

In MOND, the external field effect breaks the local symmetry between the radial and azimuthal directions. This is because the `external' gravity from the rest of the galaxy has no azimuthal component but generally does have a radial component. Thus, we expect Equation \ref{gamma} to depend on the angle $\psi$.

In general, the directional gradient operator in Equation \ref{QUMOND_g_ext_domination} is
\begin{eqnarray}
	\hw ~=~ \cos \theta \cos \psi \frac{\partial}{\partial \widetilde{r}} \, + \, \sin \theta \frac{\partial}{\partial z} \, .
\end{eqnarray}

As a result, the derivation in Sections \ref{Section_QUMOND_potential} and \ref{Boundary_condition} can be generalized to $\psi \neq 0$ by setting
\begin{eqnarray}
	\cos \theta ~\to~ \cos \theta \cos \psi \, .
\end{eqnarray}

However, $\sin \theta$ in those sections should not be altered as it represents the component of $\hw$ in the vertical direction. Consequently, Equation \ref{MOND_Phi_amplitude} becomes
\begin{eqnarray}
	A ~=~ -\frac{\mathrm{\pi} G\nu_{_0}\widetilde{\Sigma}\left( 2 + K_0 \left( \sin^2 \theta + \cos^2 \theta \cos^2 \psi \right) \right)}{\left| \alpha \right|} \, .
	\label{MOND_Phi_amplitude_spiral}
\end{eqnarray}

This leads to a generalized version of Equation \ref{gamma}.
\begin{eqnarray}
	\gamma ~=~ \nu_{_0}\left( 1 + \frac{K_0 \left( \sin^2 \theta + \cos^2 \theta \cos^2 \psi \right)}{2} \right).
	\label{gamma_psi}
\end{eqnarray}

Thus, short-wavelength non-axisymmetric disturbances in QUMOND can be analyzed similarly to Newtonian gravity if we rescale the local surface density (or equivalently the local value of $G$) by this factor. Equation \ref{gamma_psi} reduces to Equation \ref{gamma} for axisymmetric disturbances because by definition these have ${\psi = 0}$.

\section{Disk stability and surface density}
\label{Numerical_results}

Disk galaxies generally have an exponential surface density profile \citep{Freeman_1970}. In this section, we consider infinitely thin exponential disks with a range of parameters. Because MOND is an acceleration-dependent theory, the only parameter we need to vary is the central surface density $\Sigma_0$. We therefore consider a family of exponential disks with different $\Sigma_0$. Our results hold for any disk scale length $r_{_d}$ as long as the total mass is varied $\propto {r_{_d}}^2$. Otherwise, we must consider a different member of our disk family with the appropriate $\Sigma_0$.

In this and subsequent sections, we consider only the unperturbed gravitational field in disk galaxies. We therefore drop the convention that $_0$ subscripts refer to background quantities. Instead, all quantities are understood to be background values.

\subsection{Rotation curves}
\label{Rotation_curves}

The first step to finding $\Phi$ is obtaining the Newtonian potential $\PhiN$. This satisfies Equation \ref{Laplace_equation} outside the disk as there is no matter there. The presence of the disk imposes a boundary condition at all points with $z = 0$. Applying the Poisson equation to a `Gaussian pill-box' around a small part of the disk shows that $\PhiN$ must satisfy
\begin{eqnarray}
	\left. \frac{\partial \PhiN}{\partial z} \right|_{z=0^+} ~=~ 2 \pi G \Sigma ~=~ 2 \pi G \Sigma_0 \mathrm{e}^{-\frac{r}{r_{_d}}}.
\end{eqnarray}

We solve this using grid relaxation of Equation \ref{Laplace_equation}, focusing on the region $z > 0$ because of symmetry. To speed up the numerical convergence, we use successive over-relaxation on a single grid with a customized radial resolution. The details of our procedure are explained in appendix A of \citet{Banik_2018_escape}, where we also mention our initial guess for $\PhiN$, the boundary and stopping conditions and the choice of over-relaxation parameter.

Once we have found $\vgN$, we obtain $\nabla \cdot \vg$ using Equation \ref{QUMOND_basic}. The exact result will depend on the assumed MOND interpolating function, for which we use the `simple' form \citep{Milgrom_1986, Famaey_Binney_2005}. Observational reasons for preferring this form were discussed in section 7.1 of \citet{Banik_2018_Centauri}.
\begin{eqnarray}
	\nu ~=~ \frac{1}{2} \, + \, \sqrt{\frac{1}{4} + \frac{a_{_0}}{\left| \vgN \right|}} ~ .
	\label{Simple_nu}
\end{eqnarray}

We obtain $\vg$ from its divergence using direct summation done similarly to section 2.2 of \citet{Banik_2018_escape}. This exploits the axisymmetric nature of the problem using a ring library procedure. The situation is simpler here because we do not consider any external field on the galaxy.

\begin{figure}
	\centering
		\includegraphics[width = 8.5cm] {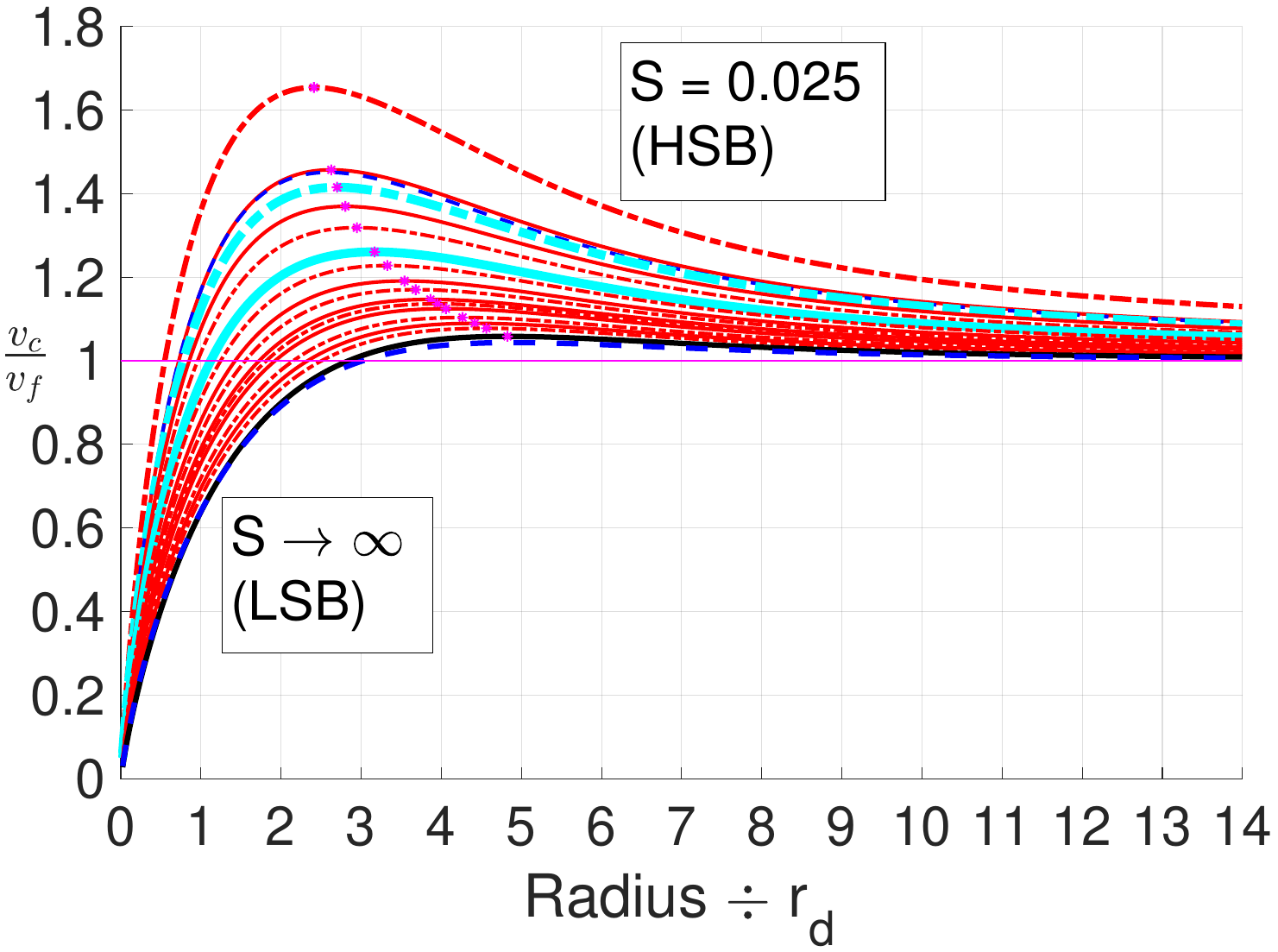}
		\caption{Rotation curves of thin exponential disk galaxies in QUMOND with different central surface density $\Sigma_0$, parameterized by $S$ in Equation \ref{S_definition}. Velocities are shown relative to the flatline level $v_{_f} = \sqrt[4]{GMa_{_0}}$ for galaxies with scale length $r_{_d}$ and total mass $M = 2\pi{r_{_d}}^2\Sigma_0$. Our adopted values for $S$ are listed in Table \ref{S_values}. The peak of each curve is indicated with a pink star. The upper cyan curve has $S = 0.06$ (similar to the Milky Way) while the lower cyan curve has $S = 0.15$ (similar to M31). To aid visibility, curves with different $S$ are shown alternating between solid and dot-dashed styles. The result for the deep-MOND limit ($S \to \infty$) is shown as a solid black curve lying below the other curves. The two thin dashed blue curves show the estimated $v_c \left( r \right)$ for $S = 0.05$ and the deep-MOND limit using the algebraic MOND relation (Equation \ref{Algebraic_MOND_approximation}). These agree very closely with the corresponding numerical curves.}
	\label{MOND_rotation_curves}
\end{figure}

We show our family of RCs in Figure \ref{MOND_rotation_curves}, normalized according to the flatline level $v_{_f}$ of each curve. The different RCs have different values of the parameter $S$, which governs the importance of MOND to the galaxy. $S$ can be thought of as the ratio between $\Sigma_0$ and the critical MOND surface density $\Sigma_M$ \citep{Milgrom_2016}.
\begin{eqnarray}
	S ~\equiv ~ \overbrace{\frac{a_{_0}}{2 \pi G}}^{\Sigma_M} \, \div \, \Sigma_0 \, .
	\label{S_definition}
\end{eqnarray}

\begin{table}
  \centering
		\begin{tabular}{ll}
			\hline
			\multicolumn{2}{c}{Values of $S$} \\
			\hline
			0.025 & 0.05 \\
			0.06 & 0.075 \\
			0.1 & 0.15 \\
			0.2 & 0.3 \\
			0.4 & 0.6 \\
			0.75 & 1 \\
			2 & 4 \\
			10 & $\infty$ \\
			\hline
		\end{tabular}
	\caption{Values of the surface density parameter $S$ (Equation \ref{S_definition}) in thin exponential disk galaxies for which we show RCs and disk stability criteria in our figures. $S \to \infty$ corresponds to the deep-MOND limit and $S \to 0$ to the Newtonian limit.}
  \label{S_values}
\end{table}

In this way, we find the outwards radial gravity $g_{_r}$ at a finite number of points along a radial transect within the disk plane. This allows us to obtain the rotation speed
\begin{eqnarray}
v_c \left( r \right) ~=~ \sqrt{-rg_{_r}} \, .
\end{eqnarray}

The curves shown in Figure \ref{MOND_rotation_curves} correspond to galaxies where $S$ takes the values listed in Table \ref{S_values}. Galaxies with lower $S$ have a higher $v_c \left( r \right)$ that flattens out at larger $r$. We alternate the line styles between solid and dashed to help identify them. The lowest black curve is the result for the deep-MOND limit ($S \to \infty$), which we obtain by using $\nu = \sqrt{\frac{\azero}{\left| \vgN \right|}}$ in Equation \ref{QUMOND_basic}.

\subsection{Minimum \texorpdfstring{$\sigma_r$}{sigma\_r}}
\label{Minimum_sigma_r}

The QUMOND RCs derived in Section \ref{Rotation_curves} allow us to estimate the minimum $\sigma_r$ required for local stability of a stellar disk via application of the Toomre stability criterion \citep{Toomre_1964}, with $G$ adjusted according to Equation \ref{Toomre_QUMOND}.
\begin{eqnarray}
	\label{Min_sigma_r_QUMOND}
	\sigma_r ~&\geq&~ \frac{3.36 \, G \nu \Sigma \left( 1 + \frac{K_0}{2} \right)}{\Omega_r}, ~~\text{ where} \\
	{\Omega_r}^2 ~&=&~ -\frac{3 g_{_r}}{r} \, - \, \frac{\partial g_{_r}}{\partial r} \, .
\end{eqnarray}

Equation \ref{Min_sigma_r_QUMOND} can be written as a constraint on the so-called Toomre $Q_*$ parameter for stellar disks \citep[][equation 6.71]{Galactic_Dynamics}.
\begin{eqnarray}
	\label{Min_Q_star_QUMOND}
	Q_* ~\equiv~ \frac{\sigma_r \Omega_r}{3.36 \, G \nu \Sigma \left( 1 + \frac{K_0}{2} \right)} ~\geq~ 1 \, .
\end{eqnarray}

In exactly the same way, we can generalize the corresponding result for isothermal gas disks with sound speed $c_s$ \citep[][equation 6.68]{Galactic_Dynamics}.
\begin{eqnarray}
	\label{Min_Q_gas_QUMOND}
	Q_{gas} ~\equiv~ \frac{c_s \Omega_r}{\mathrm{\pi} \, G \nu \Sigma \left( 1 + \frac{K_0}{2} \right)} ~\geq~ 1 \, .
\end{eqnarray}

\begin{figure}
	\centering
	\includegraphics[width = 8.5cm] {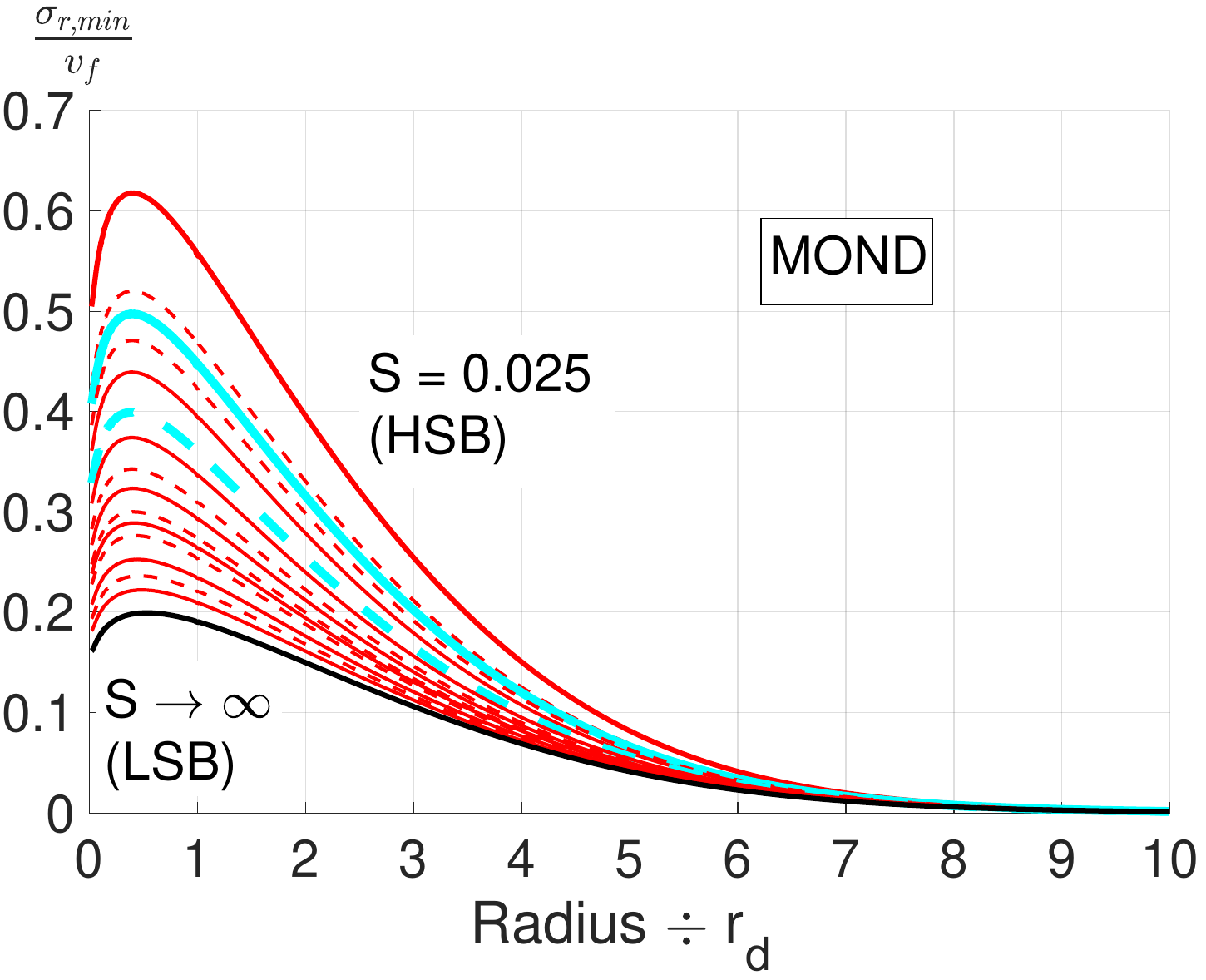}
	\caption{The minimum radial velocity dispersion $\sigma_r$ required for local stability of our family of thin exponential QUMOND disks, using the criterion in Equation \ref{Min_sigma_r_QUMOND}. We use a similar style to Figure \ref{MOND_rotation_curves} and show results for the same central surface densities (values of $S$ listed in Table \ref{S_values}).}
	\label{Min_sigma_r}
\end{figure}

Our results in Figure \ref{Min_sigma_r} show that high surface density galaxies (with low $S$) need rather high $\sigma_r$ in their central regions. Thus, MOND is unable to stabilize high surface brightness (HSB) rotationally supported disks: doing so would need such a large $\sigma_r$ that the disk would be dispersion-supported. This is in line with the fact that a low $S$ implies MOND is not very relevant to the galaxy, making it similar to a purely Newtonian exponential disk lacking a DM halo. Such systems are dynamically unstable \citep{Hohl_1971, Ostriker_Peebles_1973}.

\begin{figure}
	\centering
	\includegraphics[width = 8.5cm] {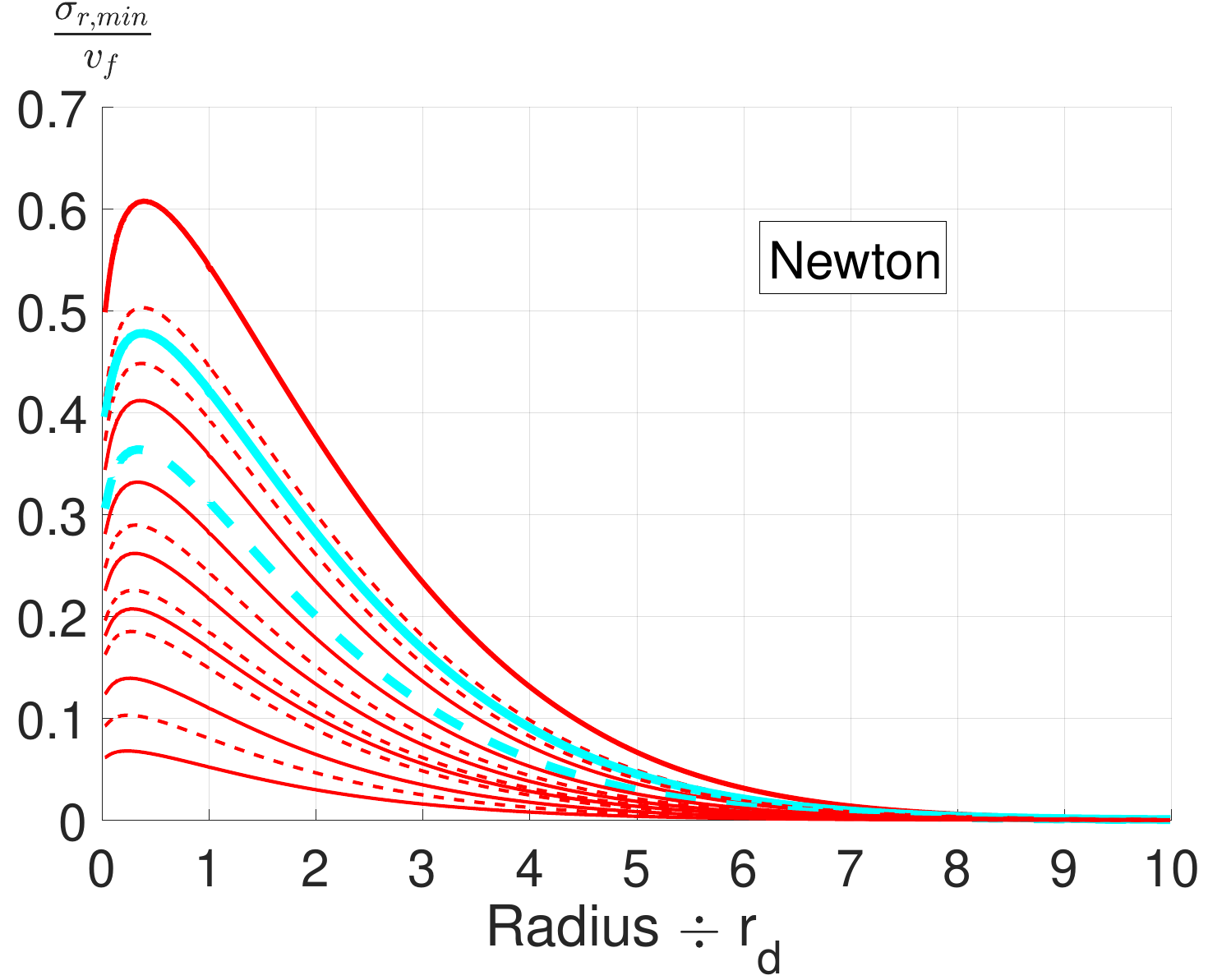}
	\caption{Similar to Figure \ref{Min_sigma_r}, but for the equivalent $\Lambda$CDM disks. These are assumed to have the same RC as the corresponding QUMOND model (Figure \ref{MOND_rotation_curves}). The stability condition follows from applying Equation \ref{Toomre_equation} to these RCs, implying no lower limit to $\sigma_r$ at very low surface density. Such disks are essentially just test particles held together by the DM halo.}
	\label{Min_sigma_r_Newton}
\end{figure}

Observationally, disk galaxies do indeed have an upper limit to their central surface brightness. Due to the difficulty of detecting LSB galaxies, the first hint of this was found by \citet{Freeman_1970}, whose figure 5 suggests that late-type disk galaxies have only a very narrow range in central surface brightness.\footnote{This was estimated by fitting an exponential profile to the outer light distribution of 36 galaxies and extrapolating the fits to the centre. This procedure works particularly well for the solid black points in his figure 5, which represent galaxies without a central cusp in the surface brightness profile.} It was later realized that the paucity of LSBs was likely a selection effect due to the brightness of the night sky \citep{Disney_1976}. The discovery of several additional LSBs eventually led to a $19\sigma$ rejection of the hypothesis that all spiral galaxies have a very narrow range of surface brightness \citep[][section 2]{McGaugh_1996}. However, higher surface brightness galaxies should be even easier to detect against the sky, suggesting that the paucity of such galaxies is a real aspect of our Universe \citep{Kruit_1987, McGaugh_1996}. The maximum surface brightness is similar to that expected in MOND for a reasonable mass to light ratio \citep{Milgrom_1989}. More recent observations confirm the presence of an upper limit \citep{Fathi_2010}.

In Figure \ref{Min_sigma_r_Newton}, we show the minimum $\sigma_r$ profiles for $\Lambda$CDM-like models in which gravity is Newtonian Newtonian but the RCs follow MOND predictions. This is because empirical RCs follow MOND expectations very closely across a huge range of galaxy surface brightness, size and mass \citep{Lelli_2017, Li_2018}. In a $\Lambda$CDM context, this is due to properties of the DM halo. Although the halo affects $v_c \left( r \right)$ and thus $\Omega_r$, it has no effect on the perturbation to $\vgN$ arising from a perturbation to $\Sigma$. Consequently, we can use the classical Toomre condition \citep{Toomre_1964} to analyze the stability of such disks, albeit with modified $\Omega_r$ for observational consistency.
\begin{eqnarray}
	\sigma_r ~\geq~ \frac{3.36 G \Sigma}{\Omega_r} \, .
	\label{Toomre_equation}
\end{eqnarray}

An interesting aspect of our results is what they reveal about disk stability in the deep-MOND limit ${S \to \infty}$ (solid black curve below the other curves). Even in this limit, a galaxy can only gain a limited amount of extra stability compared to the purely Newtonian case. To understand this, we consider how changes in $\vg$ and $\vgN$ relate to changes in the surface density $\Sigma$. In Newtonian gravity, we get that
\begin{eqnarray}
	\frac{\Delta \left| \vgN \right|}{\left| \vgN \right|} ~\approx ~ \frac{\Delta \Sigma}{\Sigma} \, .
\end{eqnarray}

In QUMOND, this is still true but $\vg \neq \vgN$ as the fields are related by Equation \ref{QUMOND_governing_equation}. Thus, we expect that
\begin{eqnarray}
	\frac{\Delta \left| \vg \right|}{\left| \vg \right|} ~&\approx &~ \frac{\Delta \left| \vgN \right|}{\left| \vgN \right|} \, + \, \frac{\Delta \nu}{\nu} \\
	~&=&~ \frac{\left( 1 + K_0 \right) \Delta \left| \vgN \right|}{\left| \vgN \right|} \, .
	\label{Perturbation_g_QUMOND}
\end{eqnarray}

Similar results would be obtained in AQUAL, whose governing equation $\nabla \cdot \left( \mu \vg \right) = -4 \mathrm{\pi} G \rho$ implies that
\begin{eqnarray}
	\frac{\left( 1 + L_0 \right) \Delta \left| \vg  \right|}{\left| \vg  \right|} ~&\approx &~ \frac{\Delta \left| \vgN \right|}{\left| \vgN \right|} \, .
	\label{Perturbation_g_AQUAL}
\end{eqnarray}

In this case, $L_0 < 1$, limiting the factor of $\left( 1 + L_0 \right)$ to at most 2. Roughly speaking, this means that self-gravitating disks in AQUAL or QUMOND can never be more than twice as stable as a bare Newtonian disk. Either modification to Newtonian gravity endows disks with a similar amount of extra stability because $\left( 1 + L_0 \right)\left( 1 + K_0 \right) \approx 1$ (Equation \ref{L0_K0_relation}).

\subsection{The critical wavelength}
\label{Section_WKB}

For our analysis to be valid, the radius $r$ must greatly exceed the wavelength $\lambda_{crit}$ most unstable to a perturbation. In this section, we explore the validity of this WKB approximation.

According to equation 22 of \citet{Toomre_1964},
\begin{eqnarray}
	\lambda_{crit} ~=~ 0.55 \times \frac{4 \pi^2 G \Sigma}{{\Omega_r}^2}
	\label{L_crit_Newton}
\end{eqnarray}
The factor of 0.55 was derived numerically in their section 5c under the assumption that $\sigma_r$ satisfies Equation \ref{Toomre_equation}. Our work shows that the appropriate MOND generalization of Equation \ref{L_crit_Newton} is
\begin{eqnarray}
	\lambda_{crit} ~=~ \frac{2.2 \pi^2 G \Sigma \nu \left( 1 + \frac{K_0}{2} \right)}{{\Omega_r}^2}
	\label{L_crit_MOND}
\end{eqnarray}

At large $r$, we expect the RC to flatline such that $\Omega_r \propto 1/r$ so that $\lambda_{crit} \propto r^2 \mathrm{e}^{-r}$. This proves that the WKB approximation should be very accurate when ${r \gg r_{_d}}$ but is likely to break down when ${r \ll r_{_d}}$.

\citet{Toomre_1964} used numerical experiments to show that the WKB approximation works quite well even when $\lambda_{crit} \approx r$ (see their section 4b). A likely explanation is that $\dvg$ at any point mostly arises from material within $\frac{1}{4}$ of a perturbation wavelength \citep{Safronov_1960}. This is due to the steep inverse square law of Newtonian gravity. Perturbations in MOND also follow an inverse square law in both the AQUAL and QUMOND formulations, though there is an additional angular dependence which is absent in Newtonian gravity \citep{Banik_2015}. Thus, we assume that our WKB approximation should work well when $\lambda_{crit} < r$.

\begin{figure}
	\centering
	\includegraphics[width = 8.5cm] {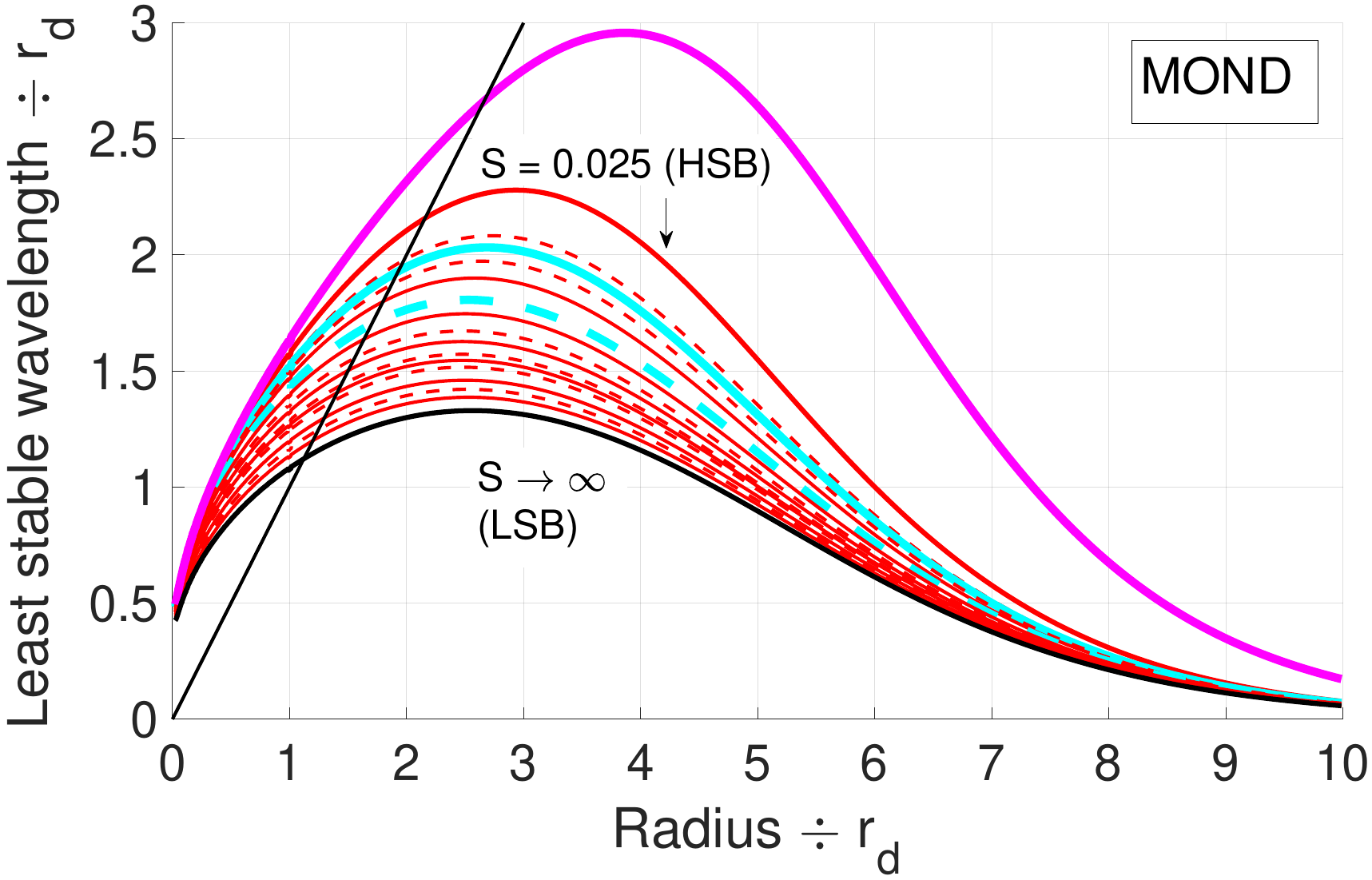}
	\includegraphics[width = 8.5cm] {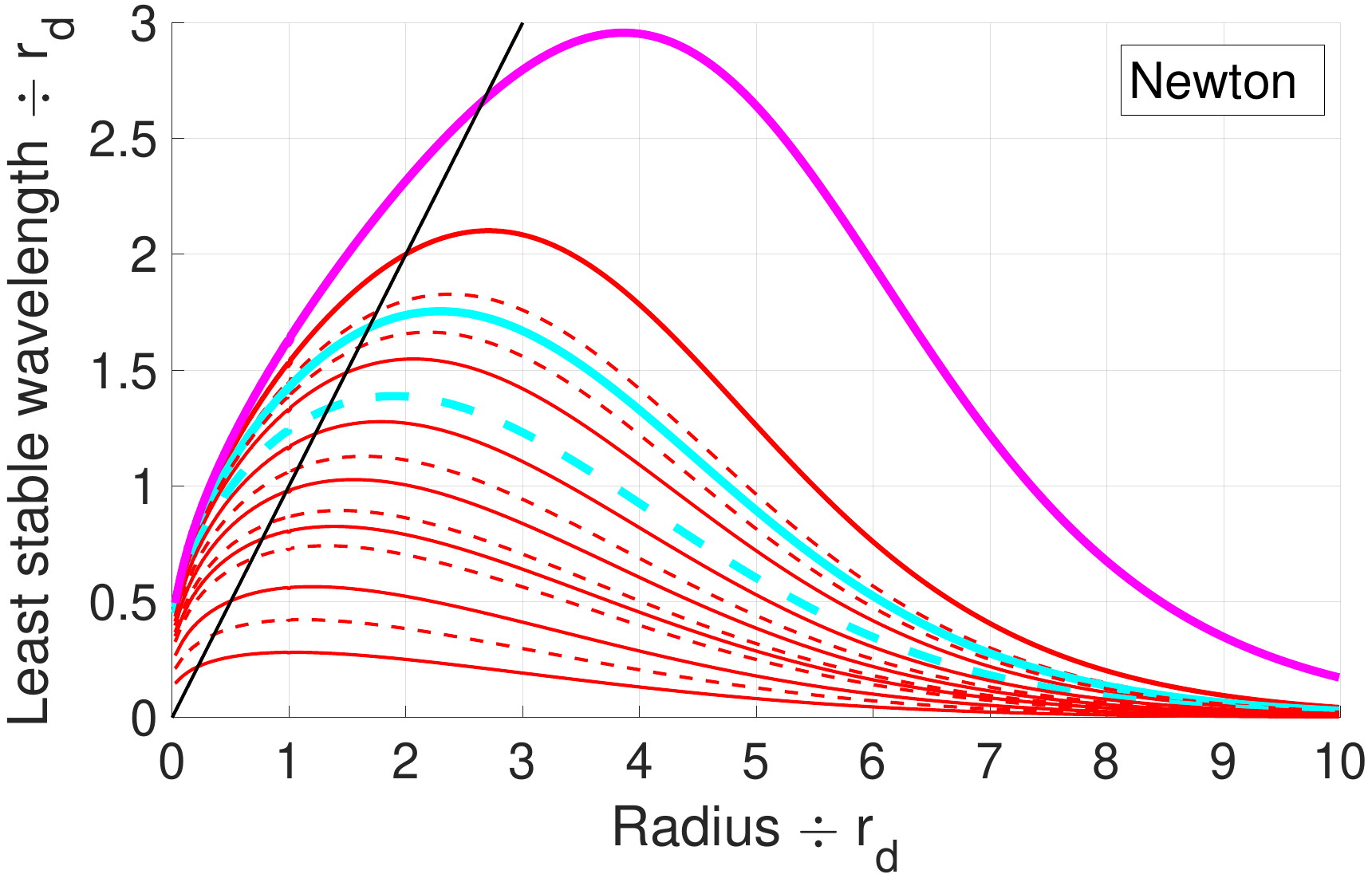}
	\caption{Our results for the wavelength most unstable to self-gravitating collapse, calculated using Equation \ref{L_crit_MOND} for MOND models (\emph{top}) and Equation \ref{L_crit_Newton} for Newtonian $\Lambda$CDM-like models (\emph{bottom}). In both panels, we show the line of equality (black) and the result for a bare Newtonian disk without any DM halo (thick pink curve at top).}
	\label{L_crit}
\end{figure}

We use Figure \ref{L_crit} to show the results of Equations \ref{L_crit_Newton} and \ref{L_crit_MOND}, each time showing a line of equality for the reasons just discussed. The bottom panel shows $\lambda_{crit}$ for $\Lambda$CDM-like models, where the RC is indistinguishable from MOND. For comparison, we also show the result for a bare Newtonian disk without any DM halo. In Newtonian models with a halo, the boost to the RC increases ${\Omega_r}$, thus reducing $\lambda_{crit}$. The reduction is more significant for a galaxy with lower $\Sigma_0$ (higher $S$) because such galaxies need a more substantial DM halo in order to explain the observed properties of LSBs.


In the MOND case, the extra factor of $\nu \left( 1 + \frac{K_0}{2} \right)$ entering into a stability analysis partially counters the boost to ${\Omega_r}$. Nonetheless, our results in the top panel of Figure \ref{L_crit} indicate that $\lambda_{crit}$ is still smaller than for a bare Newtonian disk. To understand this, suppose that the circular orbit frequency is ${\Omega_c}$. It can be straightforwardly shown that
\begin{eqnarray}
	\label{Omega_r_c_relation}
	{\Omega_r}^2 ~&=&~ \left(n + 3 \right) {\Omega_c}^2 \, ,\\
	n ~&\equiv&~ \frac{\partial \ln \left| g_r \right|}{\partial \ln r} \, .
\end{eqnarray}

Here, $n$ is the logarithmic radial derivative of $\left| g_r \right|$, the magnitude of the radial gravity. Combining Equation \ref{Omega_r_c_relation} with the fact that MOND approximately boosts $g_r$ by a factor of $\nu$ (Section \ref{Section_ALM}), we can write Equation \ref{L_crit_MOND} in terms of the Newtonian angular frequency ${\Omega_{c, N}}$.
\begin{eqnarray}
	\lambda_{crit} ~=~ \frac{2.2 \pi^2 G \Sigma \nu \left( 1 + \frac{K_0}{2} \right)}{\left(n + 3 \right) \nu {\Omega_{c, N}}^2}
	\label{L_crit_MOND_ALM}
\end{eqnarray}

The factors of $\nu$ cancel between the numerator and denominator. This is because $\nu$ enhances both the global RC and the local response of $\bm{g}$ to a density perturbation. As a result, $\lambda_{crit}$ does not differ too greatly from the Newtonian result in the absence of DM.

Even so, some differences do exist because we are left with an extra factor of $\left( 1 + \frac{K_0}{2} \right) < 1$ in the numerator of Equation \ref{L_crit_MOND_ALM}. In the deep-MOND regime, this factor is $\frac{3}{4}$. Regardless of the central $\Sigma_0$, this regime is always reached in the outskirts of a galaxy.

Some role is also played by the factor of ${\left( n + 3 \right)}$. By definition, ${n = -2}$ in the outskirts of a bare Newtonian disk. In MOND, Equation \ref{Deep_MOND_limit} implies that the analogous result is ${n = -1}$. This difference reduces $\lambda_{crit}$ by another factor of 2 compared to the Newtonian case.

Therefore, our results indicate that the WKB approximation should be even more accurate for MOND than for bare Newtonian disks of the sort analyzed by \citet{Toomre_1964}. Comparing our $\lambda_{crit}$ curves with the lines of equality in Figure \ref{L_crit}, it is clear that the least stable wavelength is much smaller than the radius beyond the central ${\approx 2 r_{_d}}$. Within this region, we expect our analysis to be less accurate. The exact details depend on $S$: the WKB approximation remains valid down to lower $r$ for galaxies with larger $S$.

This effect is much stronger for $\Lambda$CDM-like models (bottom panel of Figure \ref{L_crit}). Thus, the $\Lambda$CDM version of our analysis should be valid almost everywhere within a LSB galaxy. This is fortunate given the importance of LSBs in distinguishing between $\Lambda$CDM and MOND (Section\ \ref{Observations}).

\subsection{Comparison with algebraic MOND relations}
\label{Section_ALM}

Our results in this section have so far made use of numerical methods, even though the QUMOND version of the Toomre criterion is analytic (Equation \ref{Min_Q_star_QUMOND}). This is because QUMOND RCs are not analytic. Even so, the Newtonian RC of an exponential disk is analytic \citep{Freeman_1970}. This suggests that QUMOND RCs could be estimated analytically, paving the way for analytic estimates of disk galaxy stability in QUMOND without recourse to complicated numerical techniques. Therefore, we compare our numerically determined QUMOND RCs against two algebraic MOND relations (ALMs) between the Newtonian and MOND accelerations.

\begin{figure}
	\centering
	\includegraphics[width = 8.5cm] {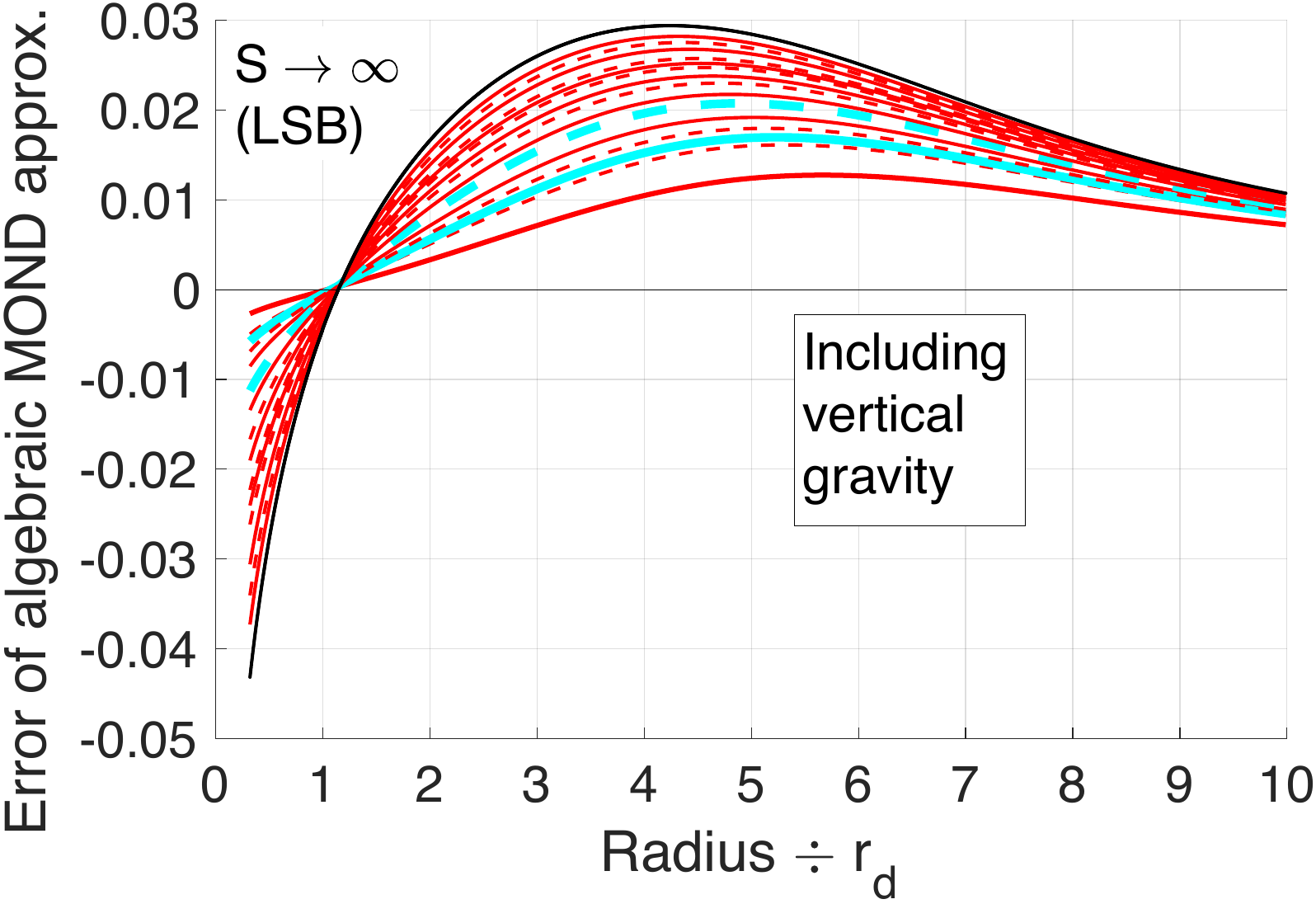}
	\caption{The fractional error $g_{_r} \div \grA - 1$ in the radial gravity $g_{_r}$ that arises from using the algebraic approximation to QUMOND (Equation \ref{Algebraic_MOND_approximation}). Results are shown down to $0.32 \, r_{_d}$ as a purely exponential profile is unlikely to remain accurate at arbitrarily low radii. A very high surface brightness galaxy ($S \to 0$) would be purely Newtonian, making the ALM exactly correct. Such a system would appear on this graph as a flat line at 0.}
	\label{ALM_error}
\end{figure}

\begin{figure}
	\centering
	\includegraphics[width = 8.5cm] {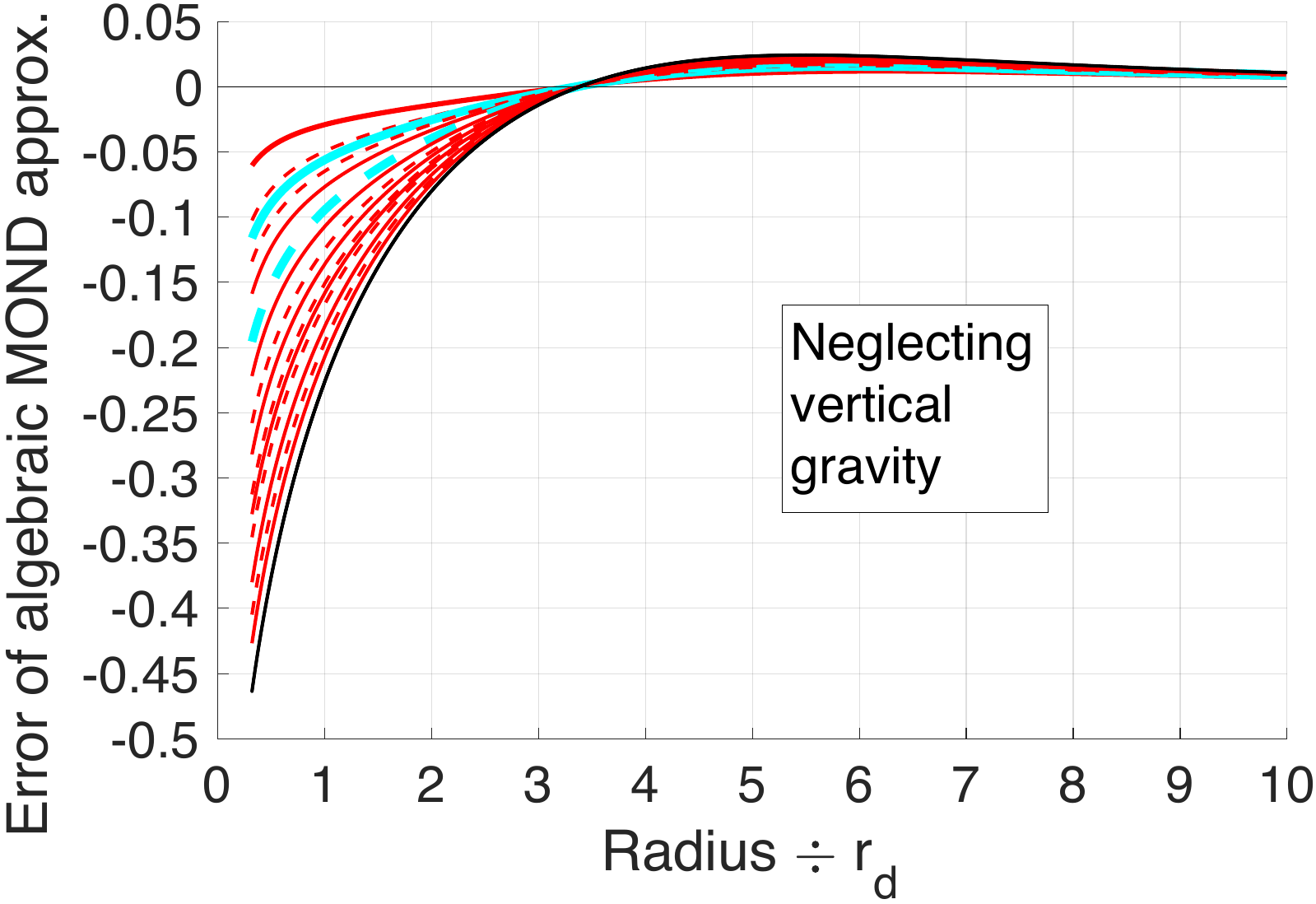}
	\caption{Similar to Figure \ref{ALM_error}, but using a different version of the ALM that neglects the vertical gravity when calculating $\nu$ (Equation \ref{MDAR}). The resulting estimates of the radial gravity now differ much more from our numerical QUMOND calculations.}
	\label{ALM_error_wrong}
\end{figure}

One such ALM relates the MOND acceleration $\grA$ to the Newtonian $\gNr$ in the disk mid-plane.
\begin{eqnarray}
	\grA ~=~ \nu \left( \frac{\gNr}{\azero} \right) \gNr \, .
	\label{MDAR}
\end{eqnarray}
This is the original formulation of MOND \citep[][equation 2]{Milgrom_1983}. When applied to RC analyses, it implies a unique relation between the accelerations predicted by Newtonian gravity and the factor by which observed accelerations exceed this prediction. This `mass discrepancy-acceleration relation' (MDAR) has been used in most subsequent MOND analyses of RCs, for example in the recent detailed analysis of \citet{Li_2018}.\footnote{They use the term `radial acceleration relation' (RAR) as the underlying cause may not be missing mass.} In modified-inertia interpretations of MOND, this equation is exactly correct for the mid-plane accelerations and can thus be used for making exact RC predictions \citep{Milgrom_1994, Milgrom_2011}.

For a modified gravity interpretation of MOND, a more appropriate ALM would equate the argument of $\nu$ with the total Newtonian acceleration just outside the disk, not the mid-plane one where ${\gNz = 0}$ \citep{Brada_1995}.
\begin{eqnarray}
	\label{Algebraic_MOND_approximation}
	\grA ~&=&~ \nu \left( \azero^{-1} \sqrt{{\gNr}^2 + {\gNz}^2} \right) \gNr, ~\text{ where} \\
	g_{_{Nz}} ~&=&~ \mp 2\pi G \Sigma \, .
\end{eqnarray}

Whichever specific version of MOND one uses, both forms of the ALM are equivalent and exact in cases of spherical symmetry, though they differ in their application to disk galaxies. For example, \citet{Brada_1995} showed in their section 3 that the ALM of Equation \ref{Algebraic_MOND_approximation} coincides exactly with AQUAL RCs for a Kuzmin disk, but the ALM of Equation \ref{MDAR} does not.\footnote{For an isolated Kuzmin disk, the gravitational field in any theory is that of a point mass offset from the galactic centre. Thus, Equation \ref{Algebraic_MOND_approximation} holds exactly everywhere outside the disk. Moreover, $g_{_r}$ is continuous vertically across the disk.}

In Figure \ref{MOND_rotation_curves}, we showed that RCs obtained with Equation \ref{Algebraic_MOND_approximation} are very similar to those based on QUMOND in the two cases considered (dashed dark blue curves), even for points quite close to the disk centre. This is similar to the results obtained by \citet{Angus_2012} and \citet{Jones_2018}.

We now consider in more detail the fractional difference in $g_{_r}$ between Equation \ref{Algebraic_MOND_approximation} and QUMOND, following on from previous calculations for AQUAL \citep{Milgrom_1986, Brada_1995}. For this purpose, we use Figure \ref{ALM_error} to show $g_{_r} \div \grA - 1$. Evidently, this ALM differs very little from QUMOND beyond the central $0.5 \, r_{_d}$, even in the deep-MOND limit. In the Newtonian limit, MOND has no effect on the dynamics, making the ALM in either form an exact representation of MOND.

The ALM of Equation \ref{MDAR} differs more significantly from QUMOND (Figure \ref{ALM_error_wrong}). Consequently, it is important to use Equation \ref{Algebraic_MOND_approximation} rather than Equation \ref{MDAR} if one is trying to approximate QUMOND with a simple algebraic relation. Even Equation \ref{Algebraic_MOND_approximation} is not exactly equivalent to QUMOND, but our results suggest that it is rather close in situations with a high degree of symmetry.

Given the increasing accuracy of observations, they may be able to distinguish between the ALMs presented here. A recent analysis comparing these ALMs with observations favored Equation \ref{Algebraic_MOND_approximation}, suggesting that MOND arises from a modification of gravity \citep{Frandsen_2018}. Some works even calculate RCs using a rigorous solution of QUMOND obtained with a Poisson solver \citep{Angus_2012, Angus_2015}. If such predictions more closely agree with observations than a simple application of Equation \ref{MDAR}, then a modified gravity interpretation of MOND would be favored. Of crucial importance to such a test would be the region ${r \approx \left(1 - 2 \right) r_{_d}}$, though accurate observations at even smaller radii would also be very helpful if non-circular motions remain small (Figure \ref{ALM_error_wrong}).

\section{Observational context}
\label{Observations}

In the $\Lambda$CDM picture, the observed internal kinematics of LSBs imply that they must be surrounded by dominant DM halos \citep[e.g.][]{McGaugh_1998}. Such inert halos would lend strong stabilizing support to LSB disks, allowing them to remain dynamically stable with a very low velocity dispersion (Figure \ref{Min_sigma_r_Newton}). Observed LSBs have rather higher velocity dispersions and thus appear to be dynamically overheated \citep{Saburova_2011}. If so, it would be difficult for them to sustain spiral density waves, the leading explanation for observed spiral features in HSBs \citep{Lin_1964}. Interestingly, LSBs also have spiral features \citep{McGaugh_1995}. It has been argued that this and other features of LSBs suggest that their gravitating mass mostly resides in their disk, contradicting $\Lambda$CDM expectations \citep[][section 3.3]{McGaugh_1998}.

Assuming the density wave theory for spiral structures, counting the number of spiral arms gives an idea of the critical wavelength most unstable to amplification by disk self-gravity. Indeed, \citet{Elena_2015} performed analytic calculations for the number of spiral arms in galaxies observed as part of the DiskMass survey \citep{Bershady_2010}. She found good agreement with observations if she made reasonable assumptions on the mass to light ratio. The DiskMass survey `selects against LSB disks', making the work of \citet{Elena_2015} an important check on the validity of the approach in a regime where $\Lambda$CDM and MOND predictions do not greatly differ.

The theory should also apply to LSBs, which provide a good opportunity to make a priori predictions because such galaxies were generally not known about in the 1960s. Our results in Section \ref{Section_WKB} show that, in a $\Lambda$CDM context, the critical wavelength for LSB galaxies is expected to be much shorter than the radius almost everywhere (bottom panel of Figure \ref{L_crit}). Thus, the WKB approximation should work particularly well in Newtonian LSB disks with massive DM halos.

Applying a similar spiral-counting technique to LSBs in a Newtonian context, \citet{Fuchs_2003} found that their disks would be much too stable to allow the formation of their observed spiral arms if their stellar masses are similar to those suggested by stellar population synthesis models \citep[e.g.][]{Bell_de_Jong_2001}. The spiral structure could be explained in $\Lambda$CDM only if much of the mass needed to explain their elevated RCs resides not in a stabilizing DM halo but rather within the disk itself. This would make the disk very massive, sometimes requiring a mass to light ratio ${>10\times}$ the Solar value in the $R$-band.

This can be understood by considering the terms in Equation \ref{Toomre_equation}. An elevated RC implies a high epicyclic frequency ${\Omega_r}$, making the observed $\sigma_r$ much higher than the minimum required for stability if we assume the disk has a conventional mass to light ratio. Physically, this arises because density perturbations in such a system would wind up quickly. This makes the disk very stable, making it difficult to form spiral arms. In Newtonian gravity, their existence implies a much higher $\Sigma$.

An unusually massive disk is also required by the Newtonian analysis of \citet{Peters_2018} to explain the pattern speeds of bars in LSBs, which are faster than expected in 3 of the 4 galaxies they considered. Similar results were obtained by \citet{Algorry_2017} upon comparing a larger sample of observed galaxies \citep{Corsini_2011, Aguerri_2015} with results from the EAGLE hydrodynamical simulations \citep{Schaye_2015, Crain_2015}. Higher numerical resolution is required to be more certain of this result.

To some extent, bars and spiral features in galaxies can be triggered by interactions with satellites \citep{Hu_2018}. However, without disk self-gravity, any spirals formed in this way would rapidly wind up and decay due to differential rotation of the disk. Thus, evidence has been mounting over several decades that the gravity in a LSB generally comes from its disk. This contradicts the $\Lambda$CDM expectation that it should mostly come from its near-spherical halo of DM given the large acceleration discrepancy at all radii in LSBs.

It has been claimed that the degree of dynamical stability observed in some galaxies appears more consistent with $\Lambda$CDM expectations than with MOND \citep{Sanchez_2016}. Their analysis found that NGC 6503 should have a bar in MOND but currently did not. In fact, this galaxy does have a faint end-on bar \citep{Kuzio_2012}. Given that bar strengths are expected to change with time, we might merely be observing its bar at a time when it is weak. Moreover, the analysis of \citet{Sanchez_2016} did not get a good MOND fit to the RC of this galaxy by assuming a distance of 5.2 Mpc and allowing a 15\% uncertainty (see their section 5.1). Recently, \citet{Li_2018} showed in their figure A1 that a good MOND fit can be obtained for a distance of 6.5 Mpc, outside this range but rather consistent with the 6.25 Mpc measured by \citet{Tully_2013}. Given the importance of the RC to the stability of a galaxy, there is clearly some doubt regarding the tension claimed by \citet{Sanchez_2016} between the MOND-predicted and observed properties of NGC 6503. At a larger heliocentric distance, the same rotation velocity implies a lower acceleration at the analogous position in the galaxy e.g. at its half-light radius. In MOND, the lack of a DM halo means this is only possible if the disk has a lower surface density. This would take it deeper into the MOND regime, endowing it with more stability (larger $S$ in Figure \ref{Min_sigma_r}).

\section{Discussion and Conclusions}
\label{Conclusions}

We consider the stability of disks to short-wavelength axisymmetric perturbations in the quasi-linear formulation of Modified Newtonian Dynamics \citep[QUMOND,][]{QUMOND}. For the same surface density perturbation, our main result is that the potential perturbation within the disk plane is enhanced compared to the Newtonian result by the factor given in Equation \ref{gamma}.

Though limited in its applicability, this result allows us to obtain the QUMOND generalization of the Toomre disk stability condition \citep{Toomre_1964}. We present this in Equation \ref{Min_Q_star_QUMOND} for stellar disks and Equation \ref{Min_Q_gas_QUMOND} for gas disks. In both cases, the radial epicyclic frequency must be based on the actual RC, which is enhanced in MOND.

We use Equation \ref{Min_Q_star_QUMOND} to estimate the minimum radial velocity dispersion $\sigma_r$ required by thin exponential disk galaxies to avoid local self-gravitating collapse. In Newtonian gravity, all such galaxies are identical up to scaling. This is no longer true in MOND as it depends non-linearly on the typical acceleration, which is fully determined by the central surface density. Thus, we set up a one-parameter family of exponential disks and numerically determine their RCs (Figure \ref{MOND_rotation_curves}) and minimum $\sigma_r$ profiles (Figure \ref{Min_sigma_r}). We also consider the stability of the analogous galaxies in $\Lambda$CDM, which we assume have a DM halo that causes their RC to match QUMOND predictions based on the baryons alone.

Numerical models show that the Toomre criterion works rather well as a stability condition for purely axisymmetric Newtonian galaxies \citep{Miller_1974}. However, systems satisfying it are generally unstable to non-axisymmetric disturbances like bars \citep{Kalnajs_1970, Kalnajs_1972}. Thus, our results should only be interpreted as a lower limit on the velocity dispersion required for disk stability.

To help future work on non-axisymmetric disturbances in MOND, we considered short-wavelength perturbations whose wave-vector lies in any direction (Section \ref{Non_axisymmetric_perturbations}). This is useful because logarithmic spiral arms have a constant pitch angle, allowing them to be locally approximated as a plane wave with normal at a fixed angle to the radial direction. In this case, Equation \ref{gamma_psi} gives the angle-dependent factor by which the potential perturbation within the disk plane is enhanced compared to the Newtonian result.

Throughout this work, we make use of the WKB approximation, namely that perturbations are much smaller than the galactocentric radius. In Section \ref{Section_WKB}, we test the validity of this approximation. Our results show that it works quite well beyond the central ${\approx 2 r_{_d}}$ of MOND exponential disks. It is expected to work particularly well for LSB disks in $\Lambda$CDM, where the WKB approximation should be very accurate almost everywhere (bottom panel of Figure \ref{L_crit}).

A global stability analysis is beyond the scope of this work because the background gravitational field can no longer be assumed constant. As a result, the global relation between surface density and potential perturbations is much more complicated than the local relation. Even so, the relation is still linear at first order and could plausibly be determined using numerical methods. This could constitute a useful extension of our work.

The very central regions of galaxies cannot be handled with our WKB approximation and may also be affected by non-axisymmetric instabilities such as bars \citep[e.g.][]{Kalnajs_1970, Sellwood_1981}. Moreover, our results pertain only to thin disks and thus do not address the possible issue of buckling instabilities \citep{Raha_1991} or the behaviour of pressure-supported regions within galaxies. In the real Universe, even disk galaxies often have a centrally concentrated bulge. We expect that our analysis applies without modification to the regions outside the bulge, once the RC is calculated appropriately and used to determine $\Omega_r$ in Equation \ref{Min_sigma_r_QUMOND}. The bulge would provide an additional source of stability in the regions outside it by enhancing the RC\footnote{and thus $\Omega_r$} but not the disk surface density. In this sense, a bulge would have a somewhat similar effect to a DM halo, though an important difference is that bulges are often directly observed while DM halos remain speculative.

Our analytic results provide a stability criterion but shed no light on how exactly instability would develop and whether non-linear effects could saturate it. Such questions need to be addressed with numerical simulations. These indicate that rotationally supported self-gravitating disks are unstable in Newtonian gravity \citep{Hohl_1971}. They are generally thought to be stabilized by a massive surrounding DM halo \citep{Ostriker_Peebles_1973}. In principle, this halo can grant an unlimited amount of stability to the disk depending on their relative masses.

The situation is very different in MOND, where the modification to gravity can only endow disks with a limited amount of extra stability. This remains true for galaxies with arbitrarily low surface density, even though MOND has a very large effect on the dynamics of such systems. Equation \ref{Perturbation_g_QUMOND} gives a rough understanding of why this is the case.

To quantify the minimum $\sigma_r$ required for Toomre stability of QUMOND exponential disks, a numerical approach is required because the RC is not analytic (Section \ref{Numerical_results}). Nonetheless, the RC can be approximated rather well analytically, as we show by comparing RCs in the action-based QUMOND with the often-used algebraic MOND expression $\vg = \nu \vgN$ (Section \ref{Section_ALM}). Our results indicate that beyond the central ${0.5 \, r_d}$, the radial force in the disk mid-plane differs by ${< 3\%}$, even if the galaxy has a very low surface density. However, this is only true if $\nu$ in the ALM expression (Equation \ref{Algebraic_MOND_approximation}) is based on both the radial and vertical components of $\vgN$, with the latter assumed to be $2 \pi G \Sigma$ as appropriate for a point just outside the disk. If this component of $\vgN$ is neglected, then similarly good agreement between the ALM and QUMOND is only attained beyond $\approx 2.5 \, r_d$ (Figure \ref{ALM_error_wrong}). Even so, this form of the ALM (Equation \ref{MDAR}) is correct in some modified inertia interpretations of MOND \citep{Milgrom_1994, Milgrom_2011}. This may provide a way to distinguish whether MOND is best understood as a modification of gravity or of inertia. A recent analysis favoured the former \citep{Frandsen_2018}.

Our analytic disk stability condition for QUMOND (Equation \ref{Min_sigma_r_QUMOND}) should prove useful when setting up stable disk galaxies in the efficient $N$-body codes Phantom of RAMSES \citep{PoR} and RAyMOND \citep{Candlish_2015} that implement QUMOND. We are currently using Phantom of RAMSES to simulate a past flyby interaction between the Milky Way and Andromeda galaxies, thus extending our work in \citet{Banik_Ryan_2018} by performing $N$-body simulations similar to those of \citet{Bilek_2018}. We hope to clarify if this flyby scenario can form structures similar to those observed in the Local Group.

\section{Acknowledgements}

IB is supported by an Alexander von Humboldt research fellowship. The authors wish to thank the referee for helping to clarify our assumptions and their range of validity. The algorithms were set up using \textsc{matlab}$^\text{\textregistered}$.

\bibliographystyle{mnras}
\bibliography{QDS_bbl}

\bsp
\label{lastpage}
\end{document}